\documentstyle[12pt]{article}
\newcommand{\beq}{\begin{equation}}
\newcommand{\eeq}{\end{equation}}
\newcommand{\bea}{\begin{eqnarray}}
\newcommand{\eea}{\end{eqnarray}}

\begin{document}
\setcounter{page}{0}
\topmargin 0pt
\oddsidemargin 5mm
\renewcommand{\thefootnote}{\fnsymbol{footnote}}
\newpage
\setcounter{page}{0}
\begin{titlepage}

\begin{flushright}
QMW-PH-95-46\\
OU-TP-96-37P\\
{\bf hep-th/9607048}\\
 July $4,~1996$
\end{flushright}
\vspace{0.5cm}
\begin{center}
{\Large {\bf Gauge Dressing of 2D Field Theories}} \\
\vspace{1.8cm}
\vspace{0.5cm}
{Ian I. Kogan$~^{1,}$\footnote{e-mail:
i.kogan1@physics.oxford.ac.uk}, Alex
Lewis$~^{1,}$\footnote{e-mail:
a.lewis1@physics.oxford.ac.uk} and  Oleg A.
Soloviev$~^{2,}$
\footnote{e-mail: O.A.Soloviev@QMW.AC.UK}} \\
\vspace{0.5cm}
{\em$~^1$Department of Physics, University of Oxford\\
1 Keble Road, Oxford, OX1 3NP, United Kingdom}\\
{\em$~^2$ Physics Department, Queen Mary and Westfield College, \\
Mile End Road, London E1 4NS, United Kingdom}\\
\vspace{0.5cm}
\renewcommand{\thefootnote}{\arabic{footnote}}
\setcounter{footnote}{0}
\begin{abstract}
{By using the gauge Ward identities, we study correlation functions
of gauged
WZNW models. We show that the gauge dressing of the correlation
functions can
be taken into account as a solution of the Knizhnik-Zamolodchikov
equation. Our
method is analogous to the analysis of the gravitational dressing of
2D field
theories.}
\end{abstract}
\vspace{0.5cm}
\centerline{July 1996}
 \end{center}
\end{titlepage}
\newpage
\section{Introduction}

There are two dimensional quantum field theories whose correlation
functions
can be computed exactly. Solvability of the given models is based
upon large
symmetries which give rise to Ward identities in the form of
differential
equations. The celebrated examples are the Wess-Zumino-Novikov-Witten
models
whose correlators obey the Knizhnik-Zamolodchikov equation
\cite{Knizhnik}.
Remarkably, it turns out that correlation functions of  CFT
coupled to 2D
gravity can be found as a solution to a differential equation
 using the Polyakov  chiral gauge approach \cite{polyakov},
 \cite{kpz} (see also \cite{chams}). This approach was used to
study  gravitationally dressed correlation functions \cite{Klebanov},
 \cite{bilal}. 

The aim of the present  letter is to  study  a  gauge  analogue 
 of the gravitational dressing  and   to analyze correlation functions of
2D field
theories gauged with respect to some Lie group. Among these theories
are coset
constructions \cite{Goddard} which play an important role in string
theory and
statistical physics. We shall concentrate on this type of 2D CFT's
which are
described as gauged WZNW models \cite{gawedski}, \cite{guadagnini},
\cite{bardacki}, \cite{Karabali} (such models were first discussed
in the early days of string theory \cite{BH}). 
A recent discussion of these 
models is \cite{witten}.

\section{Gauge dressing of the Knizhnik-Zamolodchikov equation}

A large class of 2D conformal
field theories is described by gauged WZNW models with the following
action ( we use here the same  normalization as in \cite{witten})
\begin{equation}
S(g,A)=S_{WZNW}(g)~+~{k\over2\pi}\int d^2z\mbox{Tr}\left[
Ag^{-1}\bar\partial g - \bar A\partial g g^{-1}
+Ag^{-1}\bar Ag
{}~-~
A\bar
A\right],\label{action}\end{equation}
where
\begin{equation}
S_{WZNW}(g)={k\over8\pi}\int d^2z~\mbox{Tr}g^{-1}\partial^\mu
gg^{-1}\partial_\mu g~+~{ik\over12\pi}\int
d^3z~\mbox{Tr}g^{-1}dg\wedge
g^{-1}dg\wedge g^{-1}dg\label{wznw}\end{equation}
and $g\in G$, $A,~\bar A$ are the gauge fields taking values in the
algebra
${\cal H}$ of the diagonal group of the direct product $H\times H$,
$H\in G$.

It has become usual to study gauged WZNW models with the BRST method
\cite{Karabali}. However, this method is not very much of help in
computing
correlation functions, though, in principle, the free field
realization of
these theories allows one to calculate correlators of BRST invariant
operators.
We shall pursue a different approach which is parallel to the
analysis of the
gravitational dressing of 2D field theories.

Our starting point are the equations of motion of the gauged WZNW
model:
\begin{eqnarray}
\bar\nabla(\nabla gg^{-1})&=&0,\nonumber\\ &\label{eqmotion}&\\
\bar\partial A~-~\partial\bar A~+~[A,\bar
A]&=&0,\nonumber\end{eqnarray}
where
\begin{equation}
\bar\nabla=\bar\partial~+~\bar
A,~~~~~~\nabla=\partial~+~A.\label{nabla}
\end{equation}

Under the gauge symmetry, the WZNW primary fields $\Phi_i$ and the
gauge
fields  $A,~\bar A$ transform respectively as follows
\begin{eqnarray}
\delta\Phi_i&=&\epsilon^a(t^a_i+\bar t^a_i)\Phi_i,\nonumber\\
\delta A&=&-\partial\epsilon-[\epsilon,A],\label{gtrans}\\
\delta\bar A&=&-\bar\partial\epsilon-[\epsilon,\bar
A],\nonumber\end{eqnarray}
where $t^a_i\in{\cal H}$.

In order to fix the gauge invariance, we impose the following
condition
\begin{equation}
\bar A=0.\label{newgauge}\end{equation}
The given gauge fixing gives rise to the corresponding Faddeev-Popov
ghosts
with the action
\begin{equation}
S_{ghost}=\int d^2z~\mbox{Tr}(b\partial
c).\label{ghosts}\end{equation}

In the gauge (\ref{newgauge}), the equations of motion take the
following form
\begin{eqnarray}
\bar\partial J&=&0,\nonumber\\ &\label{geqmotion}&\\
\bar\partial A&=&0,\nonumber\end{eqnarray}
where
\begin{equation}
J=-{k\over2}\partial
gg^{-1}~-~{k\over2}gAg^{-1}.\label{J}\end{equation}
Thus, $J$ is a holomorphic current in the gauge (\ref{newgauge}).
Moreover, it
has canonical commutation relations with the field $g$ and itself:
\begin{eqnarray}
\left\{J^a(w),g(z)\right\}&=&t^ag(z)\delta(w,z),\nonumber\\
&\label{canonical}&
\\
\left\{J^a(w),J^b(z)\right\}&=&f^{abc}
J^c(z)\delta(w,z)~+~k/2\delta^{a
b}\delta'(w,z).\nonumber\end{eqnarray}
The given commutators follow from the symplectic structure of the
gauged WZNW
model in the gauge (\ref{newgauge}). In this gauge, the field $A$
plays a role
of the parameter $v_0$ of the orbit of the affine group $\hat G$
\cite{Alekseev}. Therefore, the symplectic structure of the gauged
WZNW model
in the gauge (\ref{newgauge}) coincides with the symplectic structure
of the
original WZNW model \cite{Faddeev}.

There are residual symmetries which survive the gauge fixing
(\ref{newgauge}).
Under these symmetries the
fields $\Phi_i$ and the remaining gauge field $A$ transform according
to
\begin{eqnarray}
\tilde\delta\Phi_i&=&(\epsilon^A_L
t^A_i~+~\epsilon_R^a\bar t^a)\Phi_i,\nonumber\\&
\label{nonab}
\\
\tilde\delta A&=&-\partial\epsilon_R-[\epsilon_R,
A],\nonumber\end{eqnarray}
where the parameters $\epsilon_L$ and $\epsilon_R$ are
arbitrary
holomorphic functions,
\begin{equation}
\bar\partial\epsilon_{L,R}=0.
\label{tildeepsilon}\end{equation}
In eqs. (\ref{nonab}) the generators $t^A$ act on the left index of
$\Phi_i$,
whereas $\bar t^a$ act on the right index of $\Phi_i$. One can notice
that the
left residual group is extended to the whole group $G$, whereas the
right
residual group is still the subgroup $H$.

Eq. (\ref{J}) can be presented in the following form
\begin{equation}
\frac{1}{2}\partial g~+~\frac{\eta}{2}gA~+~\frac{1}{\kappa} Jg=0.
\label{maineq}\end{equation}
Here $\eta$ and $\kappa$ are some renormalization constants due to
regularization of the singular products $gA$ and $Jg$.

In order to compute $\eta$ and $\kappa$, we need to do a few things.
First of
all, we have to find how the gauge field $A$ acts on the fields
$\Phi_i$. To
this end, let us define dressed correlation functions
\begin{equation}
\langle\langle\cdot\cdot\cdot\rangle\rangle\equiv\int{\cal D}\bar
A{\cal
D}A\langle\cdot\cdot\cdot\rangle~\exp\left[-{k\over2\pi}\int
d^2z\mbox{Tr}\left\{\bar Ag^{-1}\partial g+A\bar\partial
gg^{-1}+Ag\bar Ag^{-1}~+~A\bar A\right\}\right]
,\label{dressing}\end{equation}
where $\langle\cdot\cdot\cdot\rangle$ is the correlation function
before
gauging. The latter is found as a solution to the KZ equation
\begin{equation}
\left\{
{1\over2}{\partial\over\partial z_i}~+~\sum^N_{j\ne
i}{t^A_it^A_j\over
k+c_V(G)}{1\over z_i-
z_j}\right\}\langle\Phi_1(z_1,\bar z_1)\Phi_2(z_2,\bar
z_2)\cdot\cdot\cdot\Phi_N(z_N,\bar z_N)
\rangle=0.\label{knizhnik}\end{equation}
Here $\Phi_i$ are the primary fields of the WZNW model (\ref{wznw}),
$t^A_i$
are the representations of the generators of $G$ for the fields
$\Phi_i$,
\begin{equation}
c_V={f^{abc}f^{abc}\over\dim
G}.\label{cazimirs}\end{equation}

In the gauge (\ref{newgauge}), the dressed correlation functions
(\ref{dressing}) can be presented as follows
\begin{eqnarray}
\langle\langle\Phi_1(z_1,\bar z_1)\Phi_2(z_2,\bar
z_2)\cdot\cdot\cdot\Phi_N(z_N,\bar z_N)\rangle\rangle
=\int{\cal D}b{\cal D}c\exp(-S_{ghost})~\int{\cal D}
A\exp[-S_{eff}(
A)]\nonumber\\&\label{definition} &\\
\times\int{\cal D}g~\Phi_1(z_1,\bar z_1)\Phi_2(z_2,\bar
z_2)\cdot\cdot\cdot\Phi_N(z_N,\bar
z_N)~\exp[-\Gamma(g,A)],\nonumber\end{eqnarray}
where $S_{eff}(A)$ is the effective action of the field $A$ and
$\Gamma(g,A)$
is formally identical to the original gauged WZNW action in the gauge
(\ref{newgauge}). The
action $S_{eff}$
is non-local and can be obtained by integration of the following
variation
(which follows from the Wess-Zumino anomaly condition)
\begin{equation}
\partial{\delta S_{eff}\over\delta A^a}
{}~+~f^{abc}A^c{\delta S_{eff}\over\delta A^b}
=\tau\bar\partial A^a.\label{eff}\end{equation}
Here the constant $\tau$ is to be defined from the consistency
condition
of the gauge (\ref{newgauge}), which is
\begin{equation}
J_{tot}\equiv\delta Z/\delta\bar A^a=0,~~~~~~~a=1,2,...,\dim
H,\label{cond}\end{equation}
at $\bar A=0$. Here $Z$ is the partition function of the gauged WZNW
model.
Condition (\ref{cond}) amounts to the vanishing of the central charge
of the
affine current $J_{tot}$. This in turn means that $J_{tot}$ is a
first class
constraint \cite{Karabali}. In order to use this constraint, we need
to know
the OPE of $A$ with itself. This can be derived as follows. Let us
consider the
identity
\begin{equation}
\tau\langle\langle\bar\partial A( z) A(
z_1)\cdot\cdot\cdot
A( z_N)\rangle\rangle=\int{\cal D} A~ A(
z_1)\cdot\cdot\cdot A( z_N)\left[\partial{\delta
S_{eff}\over\delta
A^a( z)}~+~f^{abc}A^c(z){\delta
S_{eff}\over\delta
A^b( z)}\right]\mbox{e}^{-S_{eff}}.\label{equality}\end{equation}
Here we used relation (\ref{eff}). Integrating by parts in the path
integral, we arrive at the following formula
\begin{eqnarray}
\tau\langle\langle A^a( z) A^{a_1}( z_1)\cdot\cdot\cdot
A^{a_N}(
z_N)\rangle\rangle
={1\over2\pi i}\sum^N_{k=1}\{{-\delta^{aa_k}\over( z-
z_k)^2}\langle\langle\ A^{a_1}( z_1)\cdot\cdot\cdot\hat
A^{a_k}_k\cdot\cdot\cdot A^{a_N}(
z_N)\rangle\rangle
\nonumber\\&\label{formula}&\\
+{f^{aa_kb}\over z-z_k}\langle\langle\ A^{a_1}( z_1)\cdot\cdot\cdot
A^b_k\cdot\cdot\cdot A^{a_N}(
z_N)\rangle\rangle\},\nonumber\end{eqnarray}
where $\hat A_k$ means that the field $A(z_k)$ is removed from the
correlator.
In the derivation of the last equation we used the following identity
\begin{equation}
\bar\partial_{\bar z}{1\over z- z_k}=2\pi
i\delta^{(2)}(z-z_k).\label{delta}\end{equation}

{}From eq. (\ref{formula}) it follows that
\begin{equation}
\tau A^a(z)A^b(0)={1\over2\pi i}\left[-{\delta^{ab}\over z^2}
{}~+~{f^{abc}\over z}A^c(0)\right]~+~\mbox{reg}.
\label{gaugeope}\end{equation}
Along with condition (\ref{cond}), the equation (\ref{gaugeope})
gives the
expression for $\tau$
\begin{equation}
\tau={i(k+2c_V(H))\over4\pi}.\label{tau}\end{equation}

We proceed to derive the Ward identity associated with the residual
symmetry
(\ref{nonab}). The Ward identity comes  from the variation of
eq.
(\ref{definition}) under transformations (\ref{nonab}). Because it
must be
zero, we obtain the following relation
\begin{eqnarray}
\sum_{k=1}^N\bar t^a_k\delta(z,z_k)\langle\langle\Phi_1(z_1,\bar
z_1)\Phi_2(z_2,\bar
z_2)\cdot\cdot\cdot\Phi_N(z_N,\bar
z_N)\rangle\rangle\nonumber\\&\label{identity} &\\
+\tau\langle\langle\bar\partial_{\bar z}A^a(z)\Phi_1(z_1,\bar
z_1)\Phi_2(z_2,\bar
z_2)\cdot\cdot\cdot\Phi_N(z_N,\bar
z_N)\rangle\rangle=0.\nonumber\end{eqnarray}
This yields
\begin{eqnarray}
2\pi\tau\langle\langle A^a(z)\Phi_1(z_1,\bar
z_1)\Phi_2(z_2,\bar
z_2)\cdot\cdot\cdot\Phi_N(z_N,\bar
z_N)\rangle\rangle\nonumber\\&\label{Ward}
&\\
=i\sum_{k=1}^N{\bar t^a_k\over z- z_k}\langle\langle\Phi_1(z_1,\bar
z_1)\Phi_2(z_2,\bar z_2)\cdot\cdot\cdot\Phi_N(z_N,\bar
z_N)\rangle\rangle,\nonumber\end{eqnarray}
which in turn gives rise to the OPE between the gauge field $A^a$ and
$\Phi_i$
\begin{equation}
 {1\over2}A^a(z)\Phi_i(0)={1\over k+2c_V(H)}{\bar t^a_i\over
z}\Phi_i(0).\label{nonabope}\end{equation}

Now we are in a position to define the product $[A^a,\Phi_i]$.
Indeed,
we can
define this according to the following rule
\begin{equation}
 A^a(z)\Phi_i(z,\bar z)=\oint{d\zeta\over2\pi i}{
A^a(\zeta)\Phi_i(z,\bar z)\over\zeta-
z},\label{product}\end{equation}
where the nominator is understood as OPE (\ref{nonabope}). Formula
(\ref{product}) is a definition of normal ordering for the product of
two
operators.

Let us come back to eq. (\ref{maineq}). Variation of
(\ref{maineq})
under the residual symmetry gives rise to the following relation
\begin{equation}
\left[1-\eta\left(1-{c_V(H)\over k+2c_V(H)}\right)\right]
\partial\epsilon_R(z)
g(z)=0.\label{anomaly}\end{equation}
{}From this relation we find the renormalization constant $\eta$
\begin{equation}
\eta={k+2c_V(H)\over k+c_V(H)}.\label{eta}\end{equation}
In the classical limit $k\to\infty$, $\eta\to1$.

With the given constant $\eta$ the equation (\ref{maineq}) reads off
\begin{equation}
\left\{{\partial\over\partial z}~+~{k+2c_V(H)\over
k+c_V(H)}A(z)~+~{2\over\kappa}J(z)\right\}g(z)=0,\label{maineq'}
\end{equation}
where $A(z)$ acts on $g$ from the right hand side.
Now the constant $\kappa$ can be calculated from the condition that
the
combination $\partial+\eta A$ acted on $g$ as a Virasoro generator
$L_{-1}$.
For this to be the case, we have to satisfy
\begin{equation}
L_{-1}g={2J^A_{-1}J^A_0\over
k+c_V(G)}g.\label{virasoro}\end{equation}

For eq. (\ref{maineq'}) to be consistent with eq.(\ref{virasoro}),
the constant
$\kappa$ has to be as follows
\begin{equation}
\kappa={1\over k+c_V(G)}.\label{kappa}\end{equation}

All in all, with the regularization given by eq. (\ref{product}) and
the
Ward identity (\ref{Ward}) the eq. (\ref{maineq'}) gives rise to the
following
differential
equation
\begin{equation}
\left\{{1\over2}
{\partial\over\partial z_i}~+~\sum^N_{j\ne i}\left({t^A_it^A_j\over
k+c_V(G)}-{\bar t^a_i\bar t^a_j\over k+c_V(H)}\right){1\over z_i-
z_j}\right\}\langle\langle\Phi_1(z_1,\bar z_1)\Phi_2(z_2,\bar
z_2)\cdot\cdot\cdot\Phi_N(z_N,\bar z_N)
\rangle\rangle=0,\label{gko}\end{equation}
where $t^A_i\in{\cal G}$ and $\bar t^a_i\in{\cal H}$. The important
comment to
be made is that there does not exist a similar equation for the
antiholomorphic
coordinate $\bar z$. This is because we use the gauge
(\ref{newgauge}) which is
not symmetrical in $z$ and $\bar z$. In this aspect, our equation
differs from
the standard KZ equation which can be written both for the holomorphic
and
antiholomorphic coordinates. However, there is Lorentz symmetry which
allows
one to partially restore the dependence on $\bar z$.

Equation (\ref{gko}) is our main result. By solving it, one can find
dressed
correlation
functions in the gauged WZNW model. The solutions can be expressed as
products of the correlation functions in the WZNW model for the group
$G$ at level $k$ and $H$ at level $-2c_V(H)-k$.
In particular, for the two-point
function
the equation yields
\begin{equation}
{1\over2}\partial\langle\langle\Phi_i(z,\bar
z)\Phi_j(0)\rangle\rangle=-\left[{t^A_it^A_j\over
k+c_V(G)}~-~{\bar t^a_i \bar t^a_j\over
k+c_V(H)}\right]{1\over z}\langle\langle\Phi_i(z,\bar
z)\Phi_j(0)\rangle\rangle.\label{two}\end{equation}
By the projective symmetry, the two-point function has the following
expression
\begin{equation}
\langle\langle\Phi_i(z,\bar
z)\Phi_j(0)\rangle\rangle={G_{ij}\over|z|^{4\Delta_i}},
\label{twopoint}
\end{equation}
where $\Delta_i$ is the anomalous conformal dimension of $\Phi_i$
after the
gauge dressing and $G_{ij}$ is the Zamolodchikov metric which can be
diagonalized. After substitution of expression (\ref{twopoint}) into
eq.
(\ref{two}), and  using the fact that, as a consequence of the
residual symmetry (\ref{nonab}), the dressed correlation functions
must be singlets of both the left residual group $G$ and the right
residual group $H$, we find
\begin{equation}
\Delta_i={c_i(G)\over k+c_V(G)}~-~{c_i(H)\over
k+c_V(H)},\label{dimensions}\end{equation}
where $c_i(G)=t^A_it^A_i,~c_i(H)=\bar t^a_i \bar t^a_i$.

\section{$SL(2)/U(1)$ coset construction}

In what follows we shall focus on the $SL(2)/U(1)$ coset
construction. This model is particularly interesting as it 
provides a description of $2D$ black holes \cite{witten2}.
 The central charge of the theory is
\begin{equation}
 c_{SL(2)/U(1)} = \frac{3k}{k+2} - 1
\end{equation}
where our convention for the sign of $k$ is opposite to  \cite{witten2}.
 So it is $k = - 9/4$ which will give the central charge $c = 26$.
 The
$U(1)$ subgroup can be either compact or non-compact. In the case
when $U(1)$
is compact, our equation takes the following form
\begin{equation}
\left\{
{\partial\over\partial z_i}~+~2\sum^N_{j\ne i}\left({t^A_it^A_j\over
k+2}-{\bar t^3_i\bar t^3_j\over k}\right){1\over z_i-
z_j}\right\}\langle\langle\Phi_1(z_1,\bar z_1)\Phi_2(z_2,\bar
z_2)\cdot\cdot\cdot\Phi_N(z_N,\bar z_N)
\rangle\rangle=0,\label{compact}
\end{equation}
where $t^A_i\in {\cal SL}(2)$. While in the non-compact case, the
equation is
\begin{equation}
\left\{
{\partial\over\partial z_i}~+~2\sum^N_{j\ne i}\left({t^A_it^A_j\over
k+2}+{\bar t^3_i\bar t^3_j\over k}\right){1\over z_i-
z_j}\right\}\langle\langle\Phi_1(z_1,\bar z_1)\Phi_2(z_2,\bar
z_2)\cdot\cdot\cdot\Phi_N(z_N,\bar z_N)
\rangle\rangle=0,\label{noncompact}
\end{equation}
The solutions for the two-point functions of fields $\Phi^{\bar m}_l$
in the representation $l$ of $SL(2)$, with $\bar t^3\Phi^{\bar m}_l
= \bar m\Phi^{\bar m}_l$ are
\begin{equation}
\langle\langle\Phi^{\bar m}_l(z,\bar
z)\Phi^{-\bar m}_l(0)\rangle\rangle={1\over|z|^{4\Delta^{\bar m}_l}},
\end{equation}
Where equation (\ref{dimensions}) becomes
\begin{equation}
\Delta^{\bar m}_l = \frac{l(l+1)}{k+2} - g_{33} \frac{\bar m^2}{k}.
\label{sl2dims} \end{equation}
In the case of a compact $U(1)$, $g_{33} = +1$, and in the
non-compact case $g_{33} = -1$. This reproduces the spectrum 
of dimensions found in \cite{dijkgraaf}.

As an example of the use of equation (\ref{maineq}) for higher
multi-point correlation functions, we consider the four-point
function for the field $g(z,\bar z)$ in the fundamental
representation of $SL(2)$ ($l=1/2$, $\bar m = \pm 1/2$). Because
of the projective invariance and the residual symmetry
(\ref{nonab}), this has the form:
\begin{equation}
\langle\langle g^{\bar m_1}_{\epsilon_1}(1)
g^{\dagger\bar m_2}_{\epsilon_2}(2)
g^{\dagger\bar m_3}_{\epsilon_3}(3)g^{\bar m_4}_{\epsilon_4}(4)
\rangle\rangle
= |(z_1-z_4)(z_2-z_3)|^{4\Delta^{1/2}_{1/2}}
\sum_{A=1,2}I_AG_A^{\bar m_1\bar m_2\bar m_3\bar m_4}(x,\bar x),
\label{4point} \end{equation}
where:
\begin{equation}
x=\frac{(z_1-z_2)(z_3-z_4)}{(z_1-z_4)(z_3-z_2)},
{}~~~~~~I_1=\delta_{\epsilon_1\epsilon_2}\delta_{\epsilon_3\epsilon_4} 
,
{}~~~~~~I_2=\delta_{\epsilon_1\epsilon_4}\delta_{\epsilon_2\epsilon_3} 
,
{}~~~~~~\bar m_1+\bar m_2+\bar m_3+\bar m_4 = 0.
\end{equation}
For the correlation function (\ref{4point}), solving equations
(\ref{compact}) and (\ref{noncompact}) gives the holomorphic
conformal blocks as products of the solutions of the KZ equations
for $SL(2)$ and $U(1)$. The general solution contains a number of
arbitrary functions of $\bar x$:
\begin {eqnarray}
G_A^{\bar m_1\bar m_2\bar m_3\bar m_4}(x,\bar x) &=&
\sum_{p=0,1}{\cal F}_{A(p)}^{\bar m_1\bar m_2\bar m_3\bar m_4}(x)
C^{\bar m_1\bar m_2\bar m_3\bar m_4}_{(p)}(\bar x) \nonumber \\
{\cal F}_{A(p)}^{\bar m_1\bar m_2\bar m_3\bar m_4}(x) &=&
F_A^{(p)}(x)f^{\bar m_1\bar m_2\bar m_3\bar m_4}(x) \nonumber \\
F_1^{(0)} (x) &=& x^{-\frac{3}{2(2+k)}} (1-x)^{\frac{1}{2(2+k)}}
F(\frac{1}{2+k}, -\frac{1}{2+k}; \frac{k}{2+k}; x) \nonumber \\
F_2^{(0)} (x) &=& \frac{1}{k} x^{\frac{1+2k}{2(2+k)}}  
(1-x)^{\frac{1}{2(2+k)}}
F(\frac{1+k}{2+k}, \frac{3+k}{2+k}; \frac{2+2k}{2+k}; x) \nonumber \\
F_1^{(1)} (x) &=& x^{\frac{1}{2(2+k)}} (1-x)^{\frac{1}{2(2+k)}}
F(\frac{1}{2+k}, \frac{3}{2+k}; \frac{4+k}{2+k}; x) \nonumber \\
F_2^{(1)} (x) &=& -2 x^{\frac{1}{2(2+k)}} (1-x)^{\frac{1}{2(2+k)}}
F(\frac{1}{2+k}, \frac{3}{2+k}; \frac{2}{2+k}; x) \nonumber \\
f^{--++}(x) = f^{++--}(x) &=& \left( \frac{x}{1-x}
\right)^{-\frac{g_{33}}{2k}}
\nonumber \\
f^{-+-+}(x) = f^{+-+-}(x) &=& \left( \frac{1-x}{x}
\right)^{-\frac{g_{33}}{2k}}
\nonumber \\
f^{-++-}(x) = f^{+--+}(x) &=& [ (x)(1-x) ]^{\frac{g_{33}}{2k}}.
\label{cblocks} \end{eqnarray}
Taking the limits as $x \to 0, 1, \infty$ gives the dimensions of
the primary fields that appear in the operator product expansions of
$g(z)g^{\dagger}(z')$ and $g(z)g(z')$. The result is $\Delta =
\Delta^0_0, \Delta^1_0, \Delta^0_1$ or $\Delta^1_1$.

Although we do not have a differential equation for the
anti-holomorphic part of the correlation function, there are a number
of conditions which it must satisfy. The full correlation function
(\ref{4point}) must be single valued in the euclidean domain
$\bar x = x^*$, and it can have no singularities except at the points
$x=0$, $x=1$ and $x=\infty$. We can also insist that the spectrum of
dimensions for the fields in the OPE should be the same for the right
and left dimensions $\bar \Delta$ and $\Delta$. Also, the full
four-point function should have the crossing symmetry
\begin{eqnarray}
G_A^{\bar m_1\bar m_2\bar m_3\bar m_4}(x,\bar x) &=&
\sum_{B=1,2}E_{AB}G_B^{\bar m_1\bar m_3\bar m_2\bar m_4}(1-x,1-\bar  
x)
\nonumber \\
&=& \sum_{B=1,2}E_{AB}G_B^{\bar m_4\bar m_2\bar m_3\bar m_1}
(1-x,1-\bar x)
\end{eqnarray}
where $E_{12}=E_{21}=1$, $E_{11}=E_{22}=0$. An ansatz which satisfies
all these requirements is:
\begin{eqnarray}
C^{--++}_{(0)}(\bar x) = C^{++--}_{(0)}(\bar x) &=& f^{++--}(\bar x)
\left\{\alpha F_1^{(0)}(\bar x) + \beta F_2^{(0)}(\bar x)\right\}
\nonumber \\
C^{--++}_{(1)}(\bar x) = C^{++--}_{(1)}(\bar x) &=& hf^{++--}(\bar x)
\left\{\alpha F_1^{(1)}(\bar x) + \beta F_2^{(1)}(\bar x)\right\}
\nonumber \\
C^{-+-+}_{(0)}(\bar x) = C^{+-+-}_{(0)}(\bar x) &=& f^{+-+-}(\bar x)
\left\{\alpha F_2^{(0)}(\bar x) + \beta F_1^{(0)}(\bar x)\right\}
\nonumber \\
C^{-+-+}_{(1)}(\bar x) = C^{+-+-}_{(1)}(\bar x) &=& hf^{+-+-}(\bar x)
\left\{\alpha F_2^{(1)}(\bar x) + \beta F_1^{(1)}(\bar x)\right\}
\nonumber \\
C^{-++-}_{(0)}(\bar x) = \delta C^{+--+}_{(0)}(\bar x) &=&
\gamma f^{+--+}(\bar x)
\left\{ F_1^{(0)}(\bar x) + F_2^{(0)}(\bar x)\right\} \nonumber \\
C^{-++-}_{(1)}(\bar x) = \delta C^{+--+}_{(1)}(\bar x) &=&
h \gamma f^{+--+}(\bar x)
\left\{F_1^{(1)}(\bar x) + F_2^{(1)}(\bar x)\right\} \nonumber \\
h &=&  \frac{1}{4} \frac{\Gamma (\frac{1}{2+k}) \Gamma
(\frac{3}{2+k})}{\Gamma (\frac{1+k}{2+k}) \Gamma (\frac{-1+k}{2+k})}
\frac{\Gamma^2 (\frac{k}{2+k})}{\Gamma^2 (\frac{2}{2+k})}.
\label{antihol} \end{eqnarray}
To determine the constants $\alpha, \beta, \gamma, \delta$, we
consider the OPE which can be deduced from equations
(\ref{cblocks}) and (\ref{antihol}). In the region close to
$x=0$, we have in particular (with $\Delta=\Delta^{1/2}_{1/2}$)
\begin{eqnarray}
G_1^{-++-}(x,\bar x) = \delta G_1^{+--+}(x,\bar x) &=&
\gamma|x|^{-4\Delta}(1 + O(x, \bar x)+ \dots) \nonumber \\
&& - h\gamma|x|^{2\Delta^0_1 - 4\Delta}(1 + O(x, \bar x)+ \dots).
\label{xto0} \end{eqnarray}
Substituting eq. (\ref{xto0}) into eq. (\ref{4point}),
we can find the OPEs:
\begin{eqnarray}
g^-_{\epsilon_1}(z, \bar z)g^{\dagger +}_{\epsilon_2}
(0) &=& \sqrt{\gamma}|z|^{-4\Delta}
(\delta_{\epsilon_1\epsilon_2}I +
2\sqrt{h}|z|^{2\Delta^0_1}
t^A_{\epsilon_1\epsilon_2}(\Phi^0_1)^A(z, \bar z) + \cdots)
\nonumber \\ \label{ope} \\
g^+_{\epsilon_1}(z, \bar z)g^{\dagger -}_{\epsilon_2}
(0) &=& \pm\sqrt{\gamma \delta}|z|^{-4\Delta}
(\delta_{\epsilon_1\epsilon_2}I +
2\epsilon\sqrt{h}|z|^{2\Delta^0_1}
t^A_{\epsilon_1\epsilon_2}(\Phi^0_1)^A(z, \bar z) + \cdots),
\nonumber \end{eqnarray}
where $\epsilon = \pm 1$.
Equations (\ref{xto0}) and (\ref{ope}) imply that we have
normalized the two-point functions as
\begin{eqnarray}
\langle\langle g^-_{\epsilon_1}(z, \bar z)
g^{\dagger +}_{\epsilon_2}(0) \rangle\rangle
&=& \sqrt{\gamma}\frac{\delta_{\epsilon_1\epsilon_2}}
{|z|^{4\Delta}}
\nonumber \\
\langle\langle g^+_{\epsilon_1}(z, \bar z)
g^{\dagger -}_{\epsilon_2}(0) \rangle\rangle
&=& \pm\sqrt{\gamma\delta}\frac{\delta_{\epsilon_1\epsilon_2}}
{|z|^{4\Delta}}
\nonumber \\
\langle\langle (\Phi^0_1)^A(z, \bar z)\
(\Phi^0_1)^B(0) \rangle\rangle &=&
\frac{\delta^{AB}}{|z|^{4\Delta^0_1}}
\end{eqnarray}
If we assume that $|\langle\langle g^-(z, \bar z)
g^{\dagger +}(0) \rangle\rangle| = |\langle\langle g^+(z, \bar z)
g^{\dagger -}(0) \rangle\rangle|$, then we must have $ \delta=1$.
Now we can substitute the OPEs (\ref{ope}) back
into the four-point function (\ref{4point}), and find the
leading terms in the expansion of $G_1^{+-+-}(x, \bar x)$.
We already have an expression for this function in equations
(\ref{cblocks}) and (\ref{antihol}), and for the two to be
consistent we must have
\begin{equation}
\beta = \pm \gamma, ~~~~~~~ \beta -2\alpha = \epsilon \beta.
\end{equation}
The only remaining ambiguity, apart from the normalization
of $g(z, \bar z)$, is in the relative sign of the two point
functions $\langle\langle g^-(z, \bar z)
g^{\dagger +}(0) \rangle\rangle$ and
$\langle\langle g^+(z, \bar z)
g^{\dagger -}(0) \rangle\rangle$, and the sign of $\epsilon$.
 The final expression
for the full four-point function is
\beq
G_A^{\bar m_1\bar m_2\bar m_3\bar m_4}(x,\bar x) =
\sum_{B=1,2}C^{\bar m_1\bar m_2\bar m_3\bar m_4}_B
|f^{\bar m_1\bar m_2\bar m_3\bar m_4}(x)|^2
\left\{F^{(0)}_A(x) F^{(0)}_B(\bar x)
+h F^{(1)}_A(x)  F^{(1)}_B(\bar x)
\right\}
\eeq
where if $\epsilon = -1$,
\begin{equation}
C^{+--+}_B=C^{-++-}_B = 1, ~~~~~~
C^{++--}_B=C^{--++}_B=C^{+-+-}_B=C^{-+-+}_B =\pm 1,
{}~~~~~~B=1,2
\end{equation}
and if $\epsilon = +1$,
\begin{eqnarray}
&C^{+--+}_B=C^{-++-}_B = 1, ~~~~~~B=1,2 \nonumber \\
&C^{++--}_2=C^{--++}_2=C^{+-+-}_1=C^{-+-+}_1 =\pm 1
\end{eqnarray}
and all other $C^{\bar m_1\bar m_2\bar m_3\bar m_4}_B$ are $0$.

\section{Conclusion}

For the gauged WZNW models, which form a particular class of 2D
gauged
theories, we have derived a differential equation which, in
principle, allows
one to find correlation functions of these models. We are
aware of an
attempt to obtain a generalized KZ equation for coset constructions
\cite{Halpern}. We think that our equation is different from the one
obtained
in \cite{Halpern}. However, we do not know what the relation is
between these
two differential equations.

 We have used our equation to show that 
the $SL(2)/U(1)$ 
 conformal blocks can be expressed as products of 
$SL(2)$ and $U(1)$ conformal blocks. We can use this 
fact to shed some light on a suggestion in \cite{kogmav} 
that logarithmic operators will arise 
 in the spectrum of $2D$ black holes.
This now implies that the logarithmic operators must also be 
present in the spectrum of the non-unitary $SL(2)$ 
WZNW model at $k=-9/4$. There are no logarithmic singularities
in the conformal blocks for fields in 
finite dimensional representations of $SL(2)$ 
(except at some special values of $k$, which do not include 
$k=-9/4$). However, in the context of $2D$ black holes, the 
spectrum includes the infinite dimensional representations 
\cite{dijkgraaf}. Unfortunately, we do not know how to 
solve the KZ equations for these infinite dimensional 
representations, but we can consider the free-field
representation for them \cite{gerasimov}. The vertex operator
in the representation $l$, $V_l$ and the ``reflected'' 
operator  $V_{-l-1}$, have the same dimension. In the case of 
$l=-1/2$, these become degenerate and we might expect the 
second operator to become a logarithmic operator. This is 
similar to the way a logarithmic operator 
(the puncture operator) appears in the Liouville model 
 \cite{bilal}, \cite{tsvelik}.
 Logarithmic operators are expected to appear in
the spectrum of a theory only when they have integer dimensions, 
which is the case for the $l=-1/2$ operator at $k=-9/4$.
The fact that we can see the appearance of logarithmic operators in
the
spectrum of $SL(2,R)/U(1)$ coset construction suggests that we can
use our
method for further systematic studying of these operators in different
models, for example non-unitary minimal models \cite{gur}.

We hope that this approach  gives us new opportunities to study
 coset models.

\section{Acknowledgments}

We have benefited from discussions with  J. Cardy, N. Mavromatos  and
 especially A. Tsvelik.

\end{document}